# RBAC Architecture Design Issues in Institutions Collaborative Environment

**M.Umar Aftab, Amna Nisar, Adeel Ashraf, Burhan Gill**

National Textile University, Faisalabad, Pakistan

ms.umaraftab@yahoo.com, nisar390@yahoo.com, adi.ashraf147@hotmail.com, gill.burhan@yahoo.com

**Dr. Asif**

Assistant Professor

National Textile University, Faisalabad, Pakistan

asif@ntu.edu.pk

## Abstract

Institutional collaborative systems focus on providing the fast, and secure connections to students, teaching and non-teaching staff members. Access control is more important in these types of systems because different kind of users access the system on different levels. So a proper architecture must be there for these kinds of systems, for providing an efficient and secure system. As lot of work was done in RBAC like for grouping, securing the system, ease of use, and for enterprise etc but no one apply all these concepts as a whole on institution level. So, this paper will be a step towards administrative load sharing, securing the system, and ease of use.

## Introduction

More secure, efficient and easy to manage access control is a big task for sensitive organizations, especially in institutional collaborative systems. Where different groups of users work jointly, share resources, communicate with each other, on common tasks. The information and resources of such kind of organizations have different kind of sensitivities like unauthorized access, information theft, efficient and secure resource sharing etc. Access control is a mechanism through which organizations can apply restrictions on resources, to protect them from unauthorized access. The access control will allow the authorized persons to access the granted resources like internet , file sharing, and read/write in a file etc or restrict the unauthorized persons from accessing the un-granted resources like server access, management of servers and access rights management etc.(ANSI 2004; Zu, Liu et al. 2009)

One of the most widely used access control model is Role Based Access Control (RBAC). RBAC is a rich technology and a great effort in the field of access control. The basic idea in RBAC is a role that is the central part of the RBAC. In normal practice the permissions are directly assigned to the users but this is the worst case because Administrator assigns or revokes permissions one by one to or from users.(Ferraiolo 2003) This process is time taking, complex and not an efficient approach towards the rights management. Previously, user management was done with groups. Permissions were assigned to groups instead of users one by one. This technique is good for user management but not for the management of permissions.(Ravi S. Sandhu, Coyne et al. 1995) In RBAC, permissions are assigned to the central layer that is role as well as users are also assigned to the roles. If administrator wants to insert or revoke a user then he/she just assigns or revokes the user to/from the role. No need to assign or revoke permissions one by one to/from the user. Through RBAC a more secure, efficient, and easy to manage network can be created.(ANSI 2004) (Ravi S. Sandhu, Coyne et al. 1995; Habib 2011)

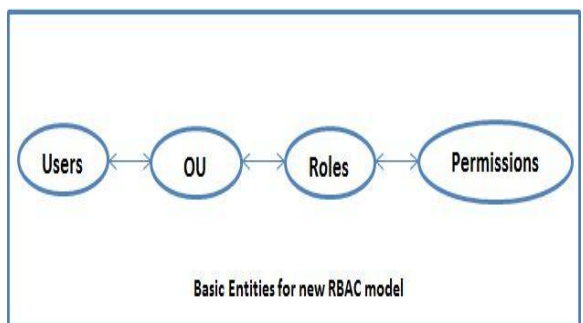

Basic Entities for new RBAC model

In this paper, we try to highlight the RBAC architectural issues in institution collaborative systems. Previously, there was no specific architecture design, defined for the institutions and more important in the context of RBAC. In this study, we proposed an architectural design in which we insert another central layer between roles and users that is **organization unit (OU)**. OU is used for directory objects contained inside domains. They are basically the containers used in active directory and we can place groups, computers, users and other OU's in it. The purpose of using OU in system is the efficient users and objects management. (Microsoft 2005) By inserting this layer the work load on the administrator will decrease and management of the access control will become easier. Secondly, we design an architecture design for institutions that is described in Hierarchical institution RBAC Architectural design portion.

## Problem Statement

We realize some problems in the existing systems. To start with the architectural design in context with RBAC in the institutions because there must be a mechanism from which an institution can run its system for the management of its IT resources.(Zu, Liu et al. 2009) On the other hand, there is no defined mechanism for managing users as the permissions like in RBAC. Number of permissions can be managed by assigning these permissions to roles. But how to manage a large number of users like students, teaching and non-teaching staff? Usually, our most concern is about the enterprise but no proper concern with the institutions.

## Methodology

RBAC is known as a marvelous model in an institution environment for access control. A role has collection of privileges in RBAC that can be assigned to users.(Ferrari 1999) As permissions are assigned to roles, which happens to be a better management of permissions thus, helping administrator in secure and easy management of the network. Before RBAC, permissions were assigned to users manually and directly without any central layer. That was not a good way towards efficient management of access control. As permission allocation and revocation was very difficult in that approach. On the other hand, organizational unit (OU) and groups' concept can be used for the management of users. Because it is really hard to manage the number of users in an organization without OU.

We proposed a system that used both the concept in a single system, RBAC with OU. The system contains an additional central layer between users and roles that is OU. By combining both concepts, we can easily manage both ends i.e. users and permissions. For managing users, we use organizational units and for managing the permissions, we use roles. The purpose of adding OU is to manage users in an efficient way, as well as decrease the extra burden of administrator. In normal practice, administrator creates, delete and manage users with the other important IT tasks like applying access control, and manage servers etc. The creation and deletion of users is a time taking job and hard to manage with other IT tasks. Sometime user deletion and creation delayed due to some other priority tasks that were not good approach towards user facilitation. We proposed two things

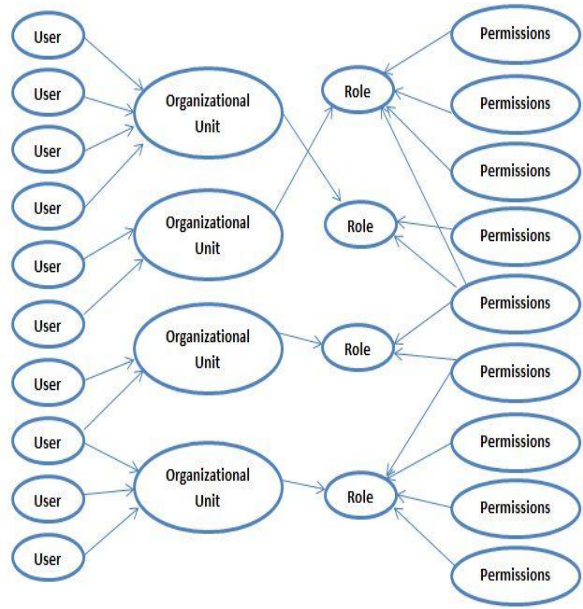

regarding user management and load sharing of administrator.

Firstly, we introduce OU concept for the efficient management of users in an institution. As institutions have number of students studying there and after every specific interval, new students are enrolled and previous students get passed from there. So, the user deletion and creation is a separate job for the institutions. By applying OU, users can be easily managed on department level and in specific department, more segregated on specific course with semester/year wise. Secondly, we proposed that the user management or OU creation will be done on department level. Because in every department there are IT coordinators that works under the instructions of administrator or IT manager. If we only assign the user creation and deletion rights to IT coordinator then the work load of administrator will be shared or decreased. As well as, this approach for load sharing is secure because IT coordinator has user creation and deletion rights only. So, these approaches will enhance user facilitation, load sharing of administrator, and efficient management of users in an institution collaborative system.

## Hierarchical Institution RBAC Architectural Design in Collaborative Environment

As discussed earlier, the paper is about the architectural design for institution RBAC in collaborative environment. As we introduce the concept of OU for the efficient management of users. So, there must be an architecture that helps the institution and IT manager for the proper deployment of RBAC in an efficient manner, with OU concept.

The Institution model is at the top, it depends upon the particular institution's structure. As well as the permission and roles, implementation and deployment depends on this model. After this the major person that will manage, control and monitor all the activities like permission, and role, assigning, deletion to OUs etc and that person is Admin RBAC Manager. He has all the rights to manage the RBAC and its instances like permissions, roles and users etc.

The working of RBAC is further segregated in three managers Permission manager, Role manager and OU manager. This segregation is more secure, efficient and a step towards load sharing. For example, if there is only one person in an institution where 10,000 plus students are studying as well as different courses are offered there. The person is the only one who will manage RBAC. Now, he is doing user creation, deletion, management of users department wise and he is applying permissions according to the requirements of departments, to roles. On the other hand, he is also managing the roles according to permissions and users. This whole scenario will create a hectic working system and this is not possible for a single person. So, this segregation and hierarchy will allow the managers to facilitate their users, system and institution in an efficient way.

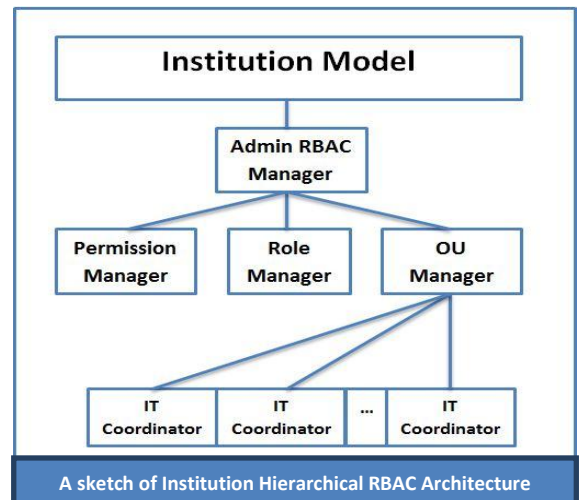

A sketch of Institution Hierarchical RBAC Architecture

The permission manager will manage, create, and revoke permission in the permission portion. He has all the rights related to permission management. The information for any kind of change, new addition, or deletion in the permissions will be given by RBAC manager.

The role manager will manage, create and revoke roles in the roles portion. He has all the rights necessary for the management of roles. As well as he will assign permissions to roles and OUs to roles. The important information like which permission must be allocated to which role or roles will be described by the RBAC manager.

The OU manager will manage, create and revoke OUs in the OU portion. He has all the rights related to OU management. He communicates with IT coordinators for the proper management of OUs. As, IT coordinators are the persons who directly interact with users like students, teaching and non-teaching staff. IT coordinators have limited rights given by the OU manager. They can create, delete or edit user information only. As well as they assign or delete users in/from the OU and they can also change the OU of any user. So, the OU manager work is just related to OU because the major work load of user management will be shared by IT coordinators. IT coordinators will pass the information of user, necessary rights for users and which rights are required for which user like internet user, system user, and thin client users etc.

Finally, the RBAC manger manages all the mangers working underneath. It manages the whole RBAC process. As well as, he guide and transfer important information to permission, role and OU managers, through which those managers start and end their working.

This whole hierarchical scenario will secure the working of RBAC in an efficient manner. Because three managers are working in parallel, after the combination of all managers work a whole RBAC scenario completes that is more secure. Like in bank lockers, a locker is open from two keys, one is of the user and second one is of the bank officer that secures the locker as compared to the single key locker. (Habib 2011) (Simon and Zurko 1997; ANSI 2004)

## Related Work

Shared and collaborative systems, multi-user apps or groupware help users or groups in cooperating and communicating for common chores. A wide range of these types of applications are available like video/audio conferences, workflow applications, shared editing/writing software and more. These systems have resources and information having different levels of sensitivity. All the applications implemented in such systems make, control, and offer access to a range of secure resources and information. Access control models are most widely used in deciding how the accessibility of information and resources in a system will be supervised and how the group decisions will be expressed. Requirements for access control in collaborative settings have been studied and stated for example where it should be applied, better scalability, must be able in protecting information, constraints for obvious access for authorized users, high degree pattern of access rights, changeable policy facility and bounded costs. Access control models for collaboration have also been discussed which include Access Matrix Model, Role Based Access, Task Based Access, Team Based Access, Spatial Access Control, Context Aware Access Control.(William 2005) Not only the significance of these models been discussed but also the shortcomings and weakness were illustrated and previously combined models were specified on the basis of preliminary work done. Evaluation criteria for these models were expressed on the basis of complexity, understandability, applicability, supporting, policy schemes, and user groups, enforcement of policies, context, and fine-grained control for particular permission.(William 2005)

Enterprise collaborative systems frequently focus on how to build practical and worthy connections amongst users, resources, information and tools. This is where, access control is helpful. Concerns and issues of enterprise access control architecture design were explained. RBAC and TBAC (Task Based Access Control) models have been examined and their benefits and shortcomings were observed in application. A better and improved R & T model (role and task-based access control model), that combines the beneficial properties of both models, has been presented. This model is general and abstract. Object-oriented method was applied to detail out the model policy schemes related to security, and to devise an organization-based task and resources or tools management method, in particular on dynamic role administration of collaborative system as a virtual organization unit (OU) where TBAC was used. A hierarchical access control architecture that can help out administrators to define, specify and implement security policy in hierarchical approach was presented. (Zu, Liu et al. 2009)

Security is an important issue, when it comes to access policies. Organizations define their own procedures/policies to prevent their information and resources from illegitimate access. Access policies or procedures are principles that state which users can perform which actions in order to implement rules of management control. Existing requirement modeling approaches have issues regarding roles and access procedures. These issues have been addressed and a framework has been proposed that will help to achieve the goals related to security. The significance of a macro-organizational had been studied and analyzed before role or actor definitions were specified in the perspective of modeling of access policies. Current modeling procedures lack this which makes it hard to clearly state the access policies and to improve them into operational constraints. A simple and new way of obtaining roles from the macro-organizational scenario had been proposed. Defining the groupings, the degrees of authority, and the management spheres/areas from which role can be described had also been confirmed. A contribution was made by demonstrating how the access policies satisfy the minimum rights and privileges. (Crook, Ince et al. 2005)

As we also gave the concept of separation of duty (SOD) in our proposed architecture. (ANSI 2004; Habib 2011) In the following paper and thesis they also work and describe the benefits of SOD. By dividing the work hierarchy, in more than one person, will secure the system. Because in the case of theft, the identification of one person is more difficult as compared to more than one person. Like in the bank example. A bank voucher contains two or more signatures from the different bank officials, just for the purpose of security and perform the SOD.(Simon and Zurko 1997)

## Conclusion

As security is the main issue for any organization. By deploying RBAC, in an organization especially in our particular domain that is institutions, will provide a more secure and efficient system for the information and resources. As well as, the deployment of RBAC with OU that will facilitate the users and administrators. In last, we design an architectural design for the RBAC based institutions that will facilitate the administrators, for the understanding the whole system working and for the proper deployment of OU-RBAC system in an institution. The proposed system will be an excellent approach towards load sharing of administrator, efficient, ease of use and the deployment of system through a proper architectural design.